\documentclass[letterpaper,11pt]{article}
\pdfoutput=1 

\usepackage{jheppub} 

\usepackage[T1]{fontenc} 

\usepackage{caption}
\usepackage{subcaption}
\usepackage{graphicx}
\usepackage{placeins}
\def\IPTD{\ensuremath{\mathrm{IP_{3D}}}}
\def\sIPTD{\ensuremath{\mathrm{sIP_{3D}}}}

\newcommand*{\TeV}{\ensuremath{\text{Te\kern -0.1em V}}}
\newcommand*{\GeV}{\ensuremath{\text{Ge\kern -0.1em V}}}
\newcommand*{\MeV}{\ensuremath{\text{Me\kern -0.1em V}}}
\newcommand*{\keV}{\ensuremath{\text{ke\kern -0.1em V}}}
\newcommand*{\eV}{\ensuremath{\text{e\kern -0.1em V}}}

\title{\boldmath A Delphes card for the EIC yellow-report detector}

\author[a,b]{Miguel Arratia}
\author[c]{Stephen Sekula,}

\affiliation[a]{Department of Physics and Astronomy, University of California, Riverside, California, 92521, USA}
\affiliation[b]{Thomas Jefferson National Accelerator Facility, Newport News, Virginia 23606, USA}
\affiliation[c]{Southern Methodist University,\\Dallas, TX, USA}

\emailAdd{miguel.arratia@ucr.edu}
\emailAdd{ssekula@smu.edu}

\abstract{The Electron-Ion Collider (EIC) Yellow Report specified parameters for the general-purpose detector that can deliver the scientific goals delineated by the EIC White Paper and NAS report. These parameters dictate the tracking momentum resolution, secondary-vertex resolutions, calorimeter energy resolutions, as well as $\pi/K/p$ ID. We have incorporated these parameters into a configuration card for Delphes, which is a widely used ``C++ framework, for performing a fast multipurpose detector response simulation''. We include both the 1.5 T and 3.0 T scenarios. We also show the expected performance for high-level quantities such as jets, missing transverse energy, charm tagging, and others. These parametrizations can be easily updated with more refined Geant4 studies, which provides an efficient way to perform simulations to benchmark a variety of observables using state-of-the art event generators such as Pythia8. }

\begin{document} 
\maketitle
\flushbottom

\section{Introduction}
\label{sec:intro}

We implement a model of the Electron-Ion Collider (EIC) reference detector described in the EIC Yellow Report~\cite{Khalek2021} into Delphes~\cite{deFavereau:2013fsa}, which is a ``C++ framework, for performing a fast multipurpose detector response simulation''. The Delphes package is widely used in studies of future collider facilities such as: CLIC, HL-LHC, FCC, ILC and Muon Collider. While it is not yet in widespread use in the EIC community, its adoption has increased significantly recently. For example, it was used in several studies included in the EIC Yellow Report (event shapes, charm tagging, jet and hadronic final-state reconstruction), as well as various publications (the most recent being a study of ``Charged Lepton-Flavor Violation at the EIC"~\cite{Cirigliano:2021img}). 

We expect that with the current call for EIC detector proposals and imminent forming of collaborations, Delphes usage in the EIC community will increase as it provides an excellent platform to complement Geant4-based studies. Among its advantages are the treatment of high-level objects such as isolated electrons, jets, taus, and missing energy; the implementation of the energy-flow algorithm; the propagation of charged particles in the solenoidal field; particle-ID matrices; and photon conversions into electron-positron pairs for a given material budget. Other alternative frameworks used  to perform EIC fast simulations (perhaps just for historical reasons) do not come even close to having these features, which have been developed over many years and tested by hundreds of active users. We argue that the ongoing formation of EIC collaborations provides an excellent opportunity to start afresh and embrace Delphes. 

\begin{figure}
    \centering
    \includegraphics[width=0.49\textwidth]{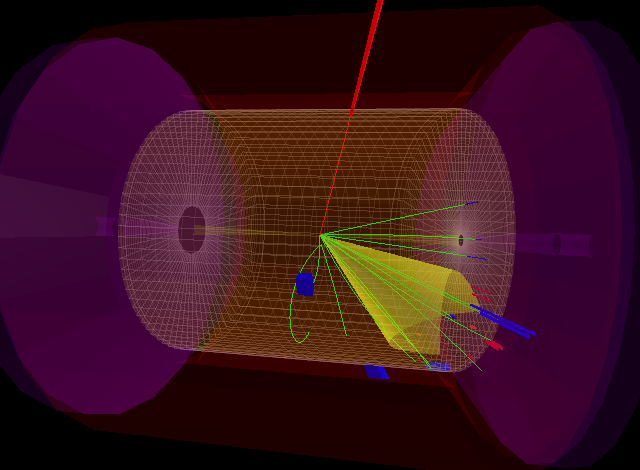}
    \includegraphics[width=0.49\textwidth]{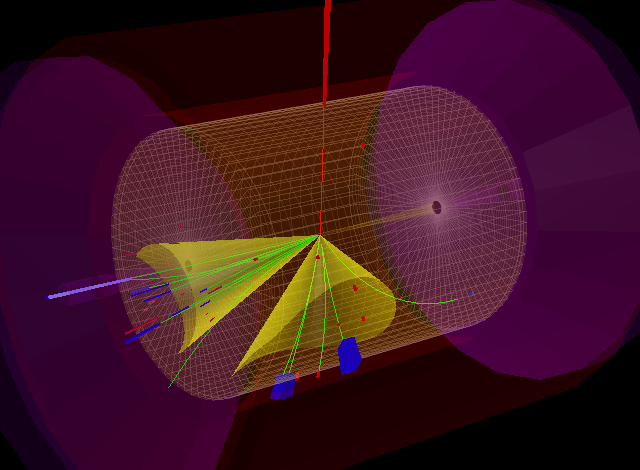}\\
        \includegraphics[width=0.49\textwidth]{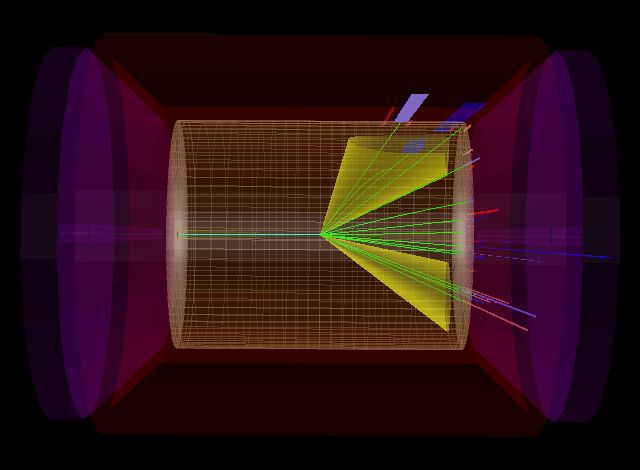}
     \includegraphics[width=0.49\textwidth]{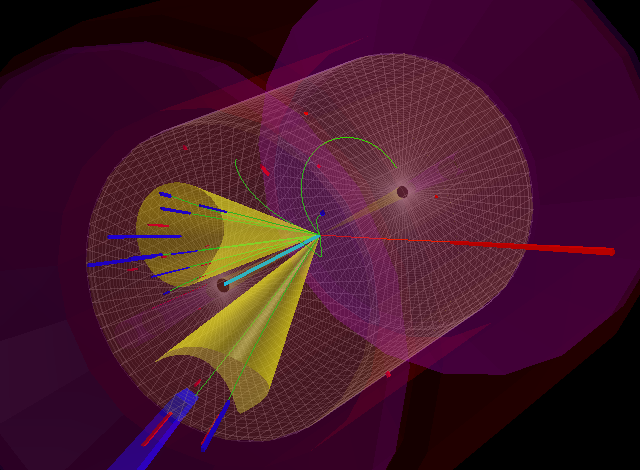}

    \caption{Event displays for simulated EIC events. Single-jet in from a Born-level DIS event (upper left); di-jet in a photon-gluon fusion event (upper right); di-jet in a photo-production event (bottom left); di-jet in a hard-diffractive DIS event (bottom right). }
    \label{fig:eventdisplays}
\end{figure}

We use the EIC Delphes card to estimate the expected performance for jets, missing transverse energy, charm-tagging, and PID. We simulate electron-proton collisions using Pythia 8.3~\cite{Sjostrand:2014zea} using a 10~\GeV\ electron beam and a 275~\GeV\ proton beam ($\sqrt{s}=105~\GeV$), which is the beam configuration that maximizes luminosity in the nominal EIC design. Some examples of event displays made with Delphes are shown in figure~\ref{fig:eventdisplays}. The Delphes card is available for use in other studies and can be downloaded from~\cite{Arratia:2021ms}. Here we show results for the parameters that correspond to the design with a 3.0 T solenoidal field.

\section{Jet Resolution}
\begin{figure}
    \centering
    \includegraphics[width=0.49\textwidth]{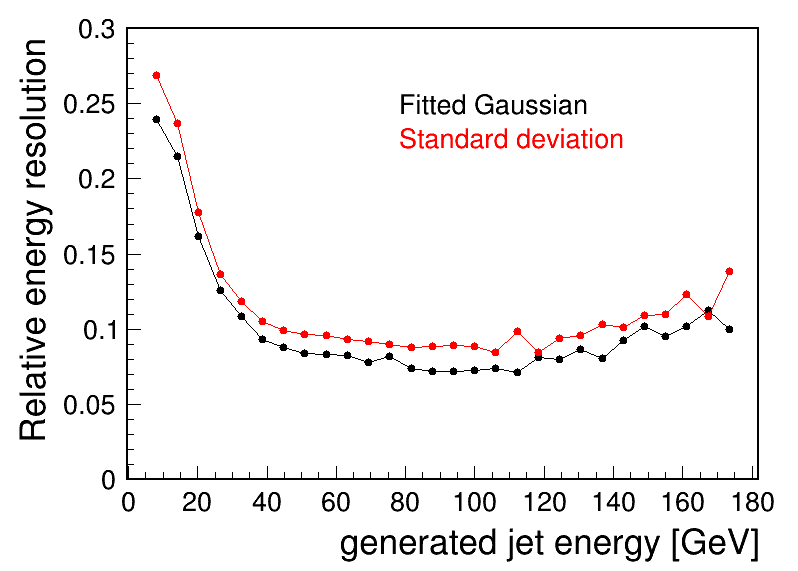}
    \includegraphics[width=0.49\textwidth]{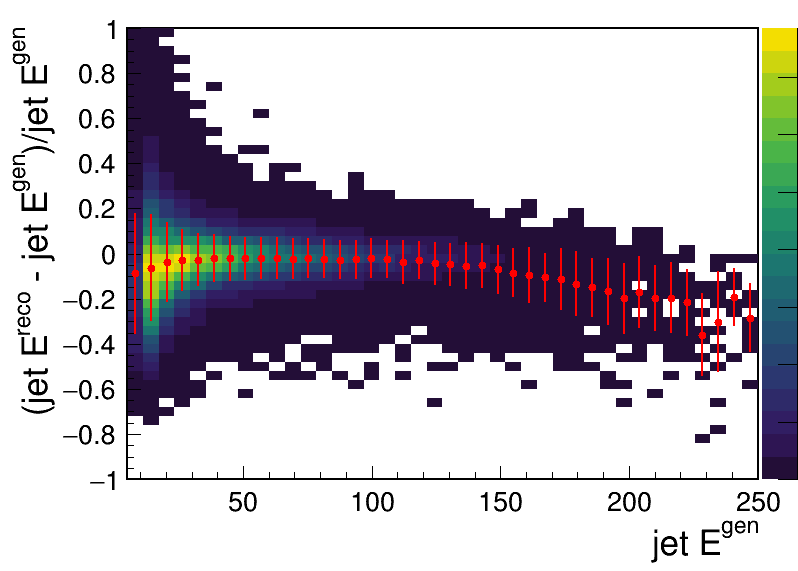}
    \caption{Jet-energy resolution and scale. The jets are defined with the anti-$k_{T}$ algorithm and $R=1.0$. The jets are reconstructed with the energy-flow algorithm as implemented in Delphes. }
    \label{fig:jetperformance}
\end{figure}

We evaluate the jet performance for jets reconstructed with the anti-$k_{T}$ algorithm and $R=1.0$. The input for the jet reconstruction are energy-flow objects as defined in Delphes. While the energy-flow algorithm implemented in Delphes is rather simplified, it has been shown that it can reproduce the  performance of the CMS experiment reasonably well~\cite{deFavereau:2013fsa}. Figure~\ref{fig:jetperformance} shows the jet-energy resolution as well as jet-energy scale obtained in neutral-current DIS events.  

\section{Missing Transverse-Energy Resolution}
We estimate the performance of reconstruction of missing transverse energy (MET) using charged-current DIS events (which yield an energetic neutrino in the final state). The MET performance is also relevant for the reconstruction of DIS kinematic variables using the Jacquet-Blondel method, which yields better performance than the lepton method in certain regions of phase-space in neutral-current DIS.

The MET is defined as the magnitude of the vector sum of the transverse momenta of all Delphes energy-flow objects. At the generator level, it is defined in a similar way but using all stable and detectable particles.

\begin{figure}
    \centering
    \includegraphics[width=0.49\textwidth]{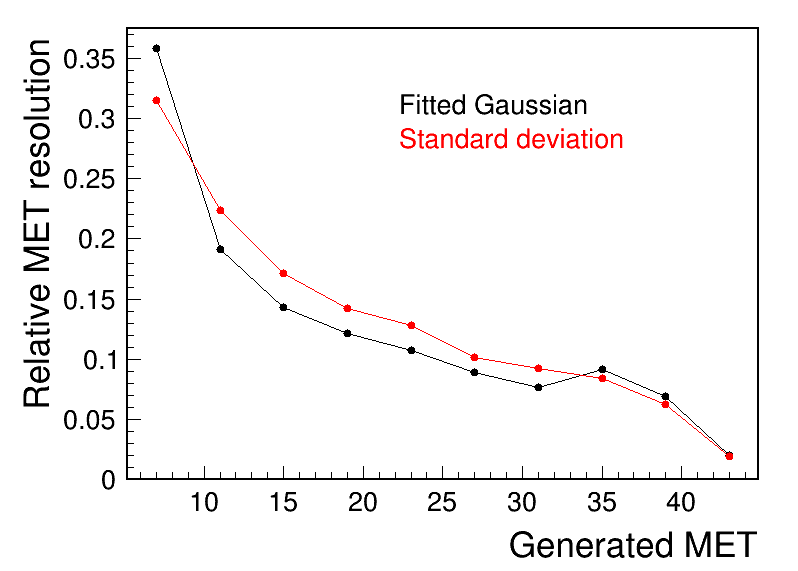}
    \includegraphics[width=0.49\textwidth]{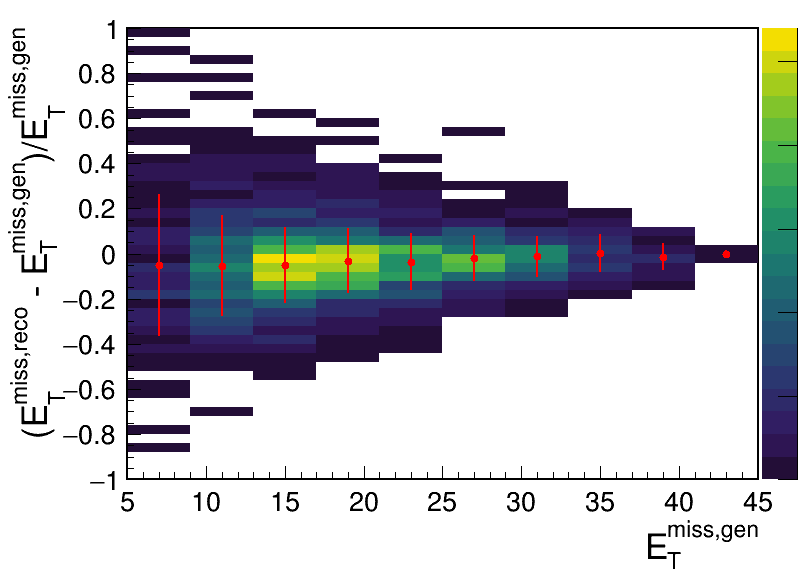}
    \caption{Missing transverse energy (MET) resolution and scale. The MET is defined with the vector sum of all energy-flow objects in the event. }
    \label{fig:metperformance}
\end{figure}
\section{Flavor-Tagging Performance}

\begin{figure}
    \centering
     \begin{subfigure}[b]{0.475\textwidth}
         \centering
         \includegraphics[width=\textwidth]{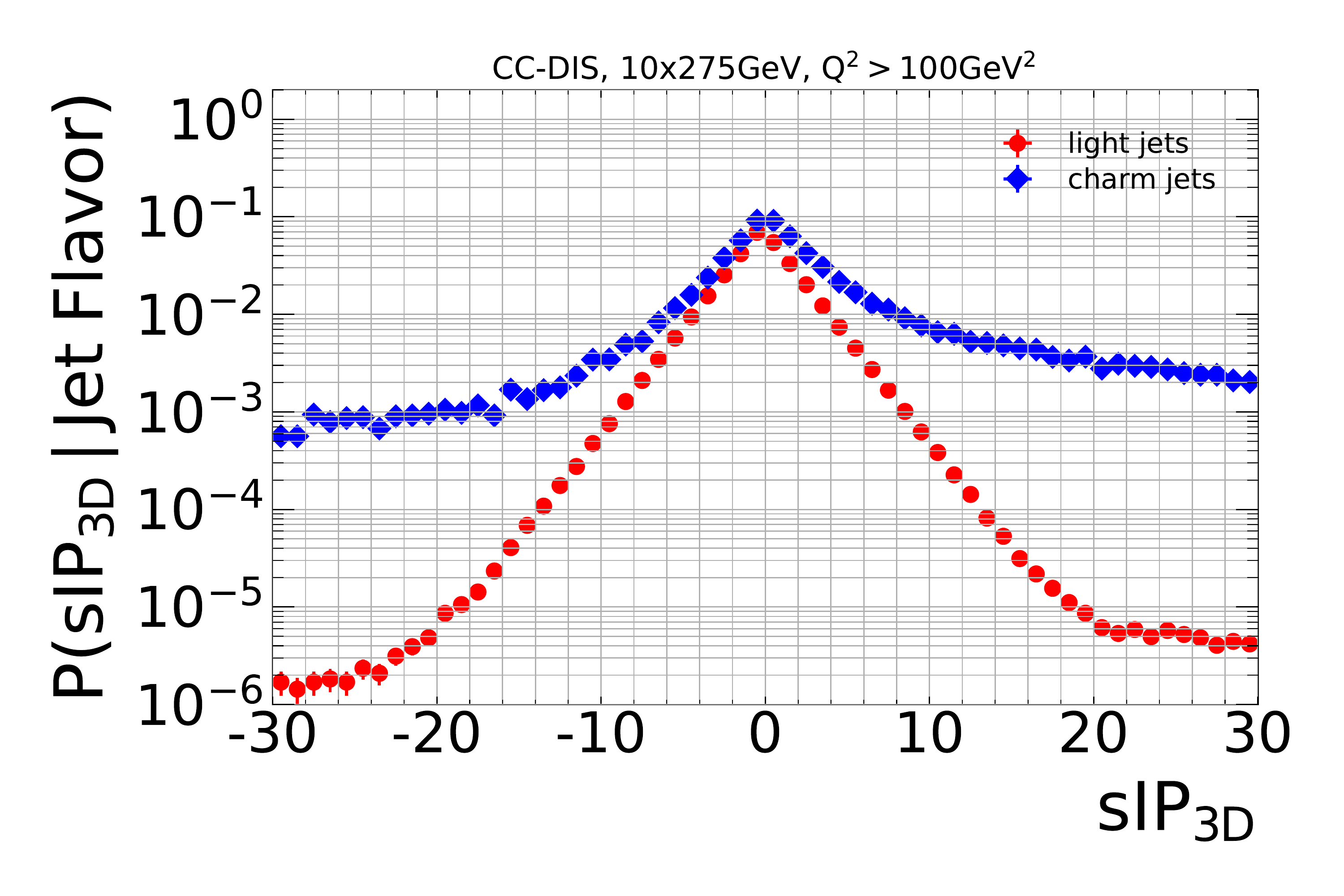}
         \caption{Probability vs.~$\sIPTD{}$ for tracks matched to light-flavor or charm jets.} 
         \label{fig:track_ip}
     \end{subfigure}
     \hfill
     \begin{subfigure}[b]{0.475\textwidth}
         \centering
         \includegraphics[width=\textwidth]{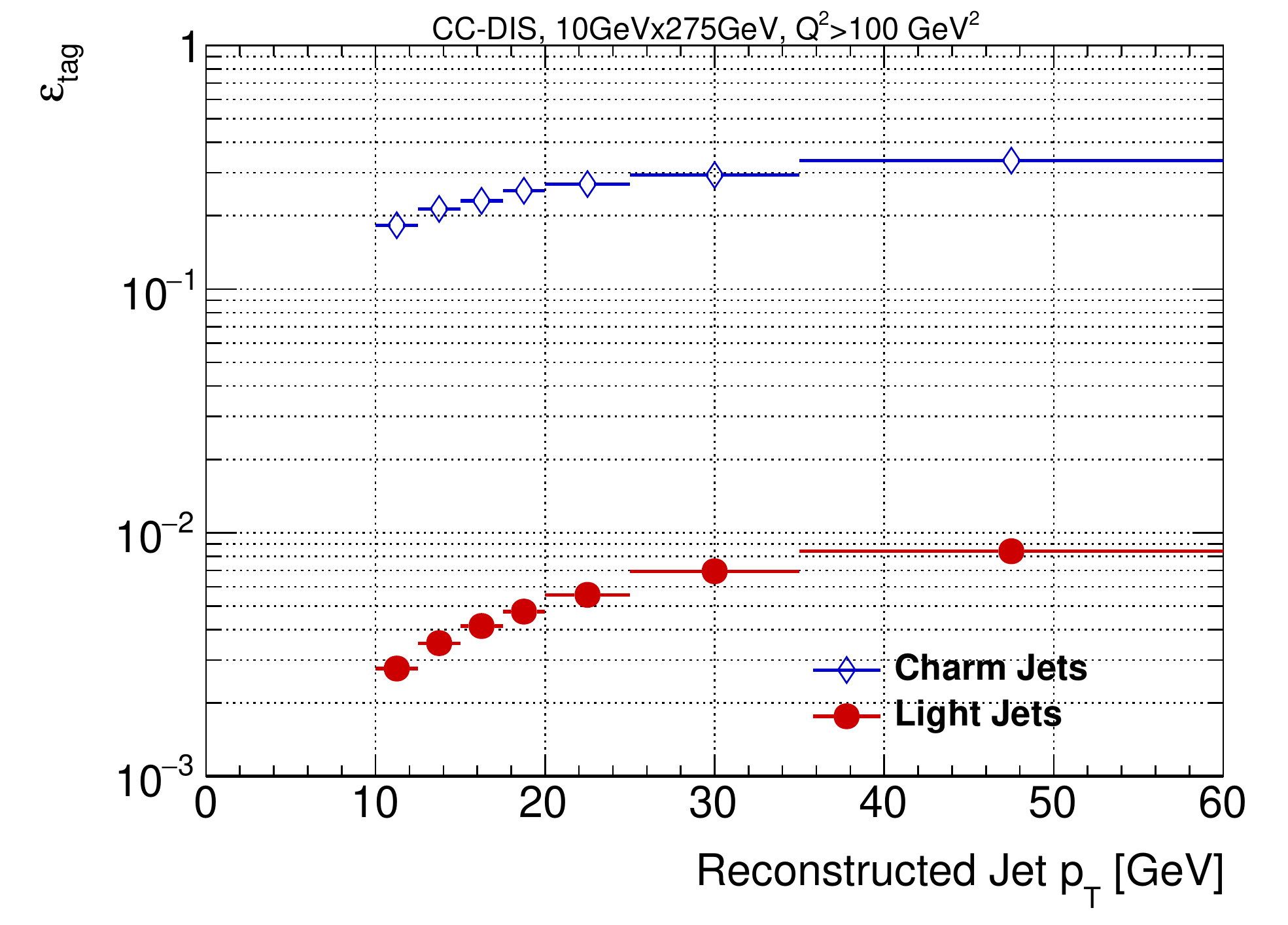}
         \caption{Displaced track-counting flavour tagging efficiency for light-flavor and charm jets.} 
         \label{fig:jet_efficiency}
     \end{subfigure}
    
    \caption{A simulation of CC DIS events is used to obtain the signed 3-D track impact parameter (left) and overall displaced track-counting flavor-tagging efficiency (right).}
    \label{fig:flavor_tagging}
\end{figure}

We utilize a displaced track-counting approach as in Ref.~\cite{Arratia:2020azl}. Tracks are first associated with jets. The impact parameter of the track (point of closest approach to the interaction point) is determined in the transverse ($d_0$) and longitudinal ($z_0$) directions, along with associated uncertainties ($\sigma_{d_0}$ and $\sigma_{z_0}$) set in the Delphes card. The 3-D impact parameter significance is then defined as $\IPTD=\sqrt{(d_0/\sigma_{d_0})^2+(z_0/\sigma_{z_0})^2}$. The signed impact parameter, $\sIPTD$ (Fig.~\ref{fig:track_ip}), is obtained by multiplying $\IPTD$ with the sign of the product $\vec{p}_j \cdot \vec{r}_{\mathrm{track}}$, where $\vec{p}_j$ is the parent jet momentum and $\vec{r}_{\mathrm{track}}$ is a vector that points from the interaction point to the point of closest approach on the track. 

A simple displaced track-counting approach is then used to flavor-tag the jets. A jet is labeled as a "charm jet" if it contains tracks that meet the following criteria: the jet contains $\ge 2$ tracks, each of which satisfies $p_T^{track}>0.5\mathrm{~GeV}$; $\sIPTD>3$; and $\sqrt{d_0^2 + z_0^2}<3\mathrm{mm}$. The efficiency for light-flavor and charm jets is shown in Fig.~\ref{fig:jet_efficiency}.

\section{Particle Identification Performance}

\begin{figure}
    \centering
     \begin{subfigure}[b]{0.475\textwidth}
         \centering
         \includegraphics[width=\textwidth]{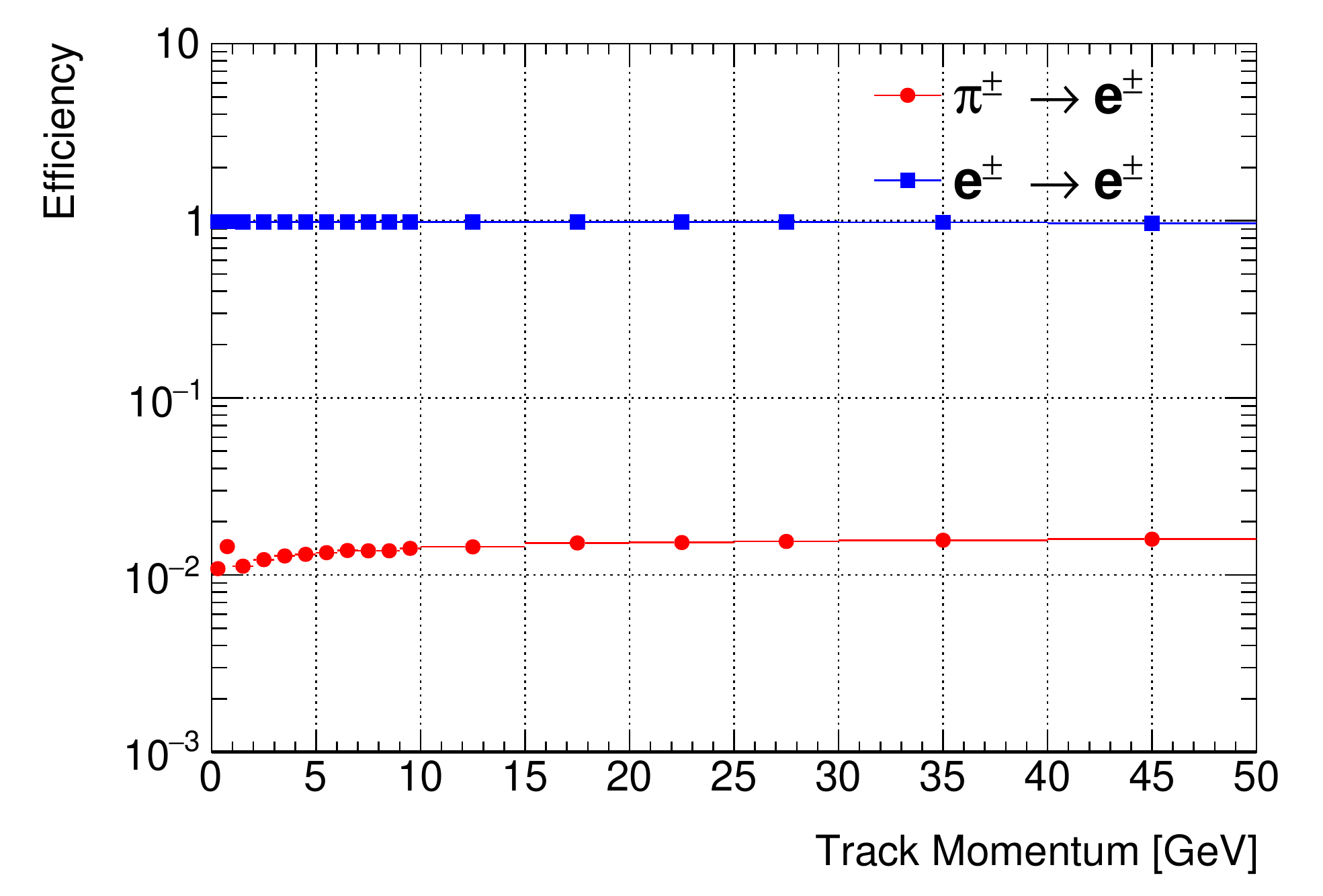}
         \caption{Electron Identification.} 
         \label{fig:eid}
     \end{subfigure}
     \hfill
     \begin{subfigure}[b]{0.475\textwidth}
         \centering
         \includegraphics[width=\textwidth]{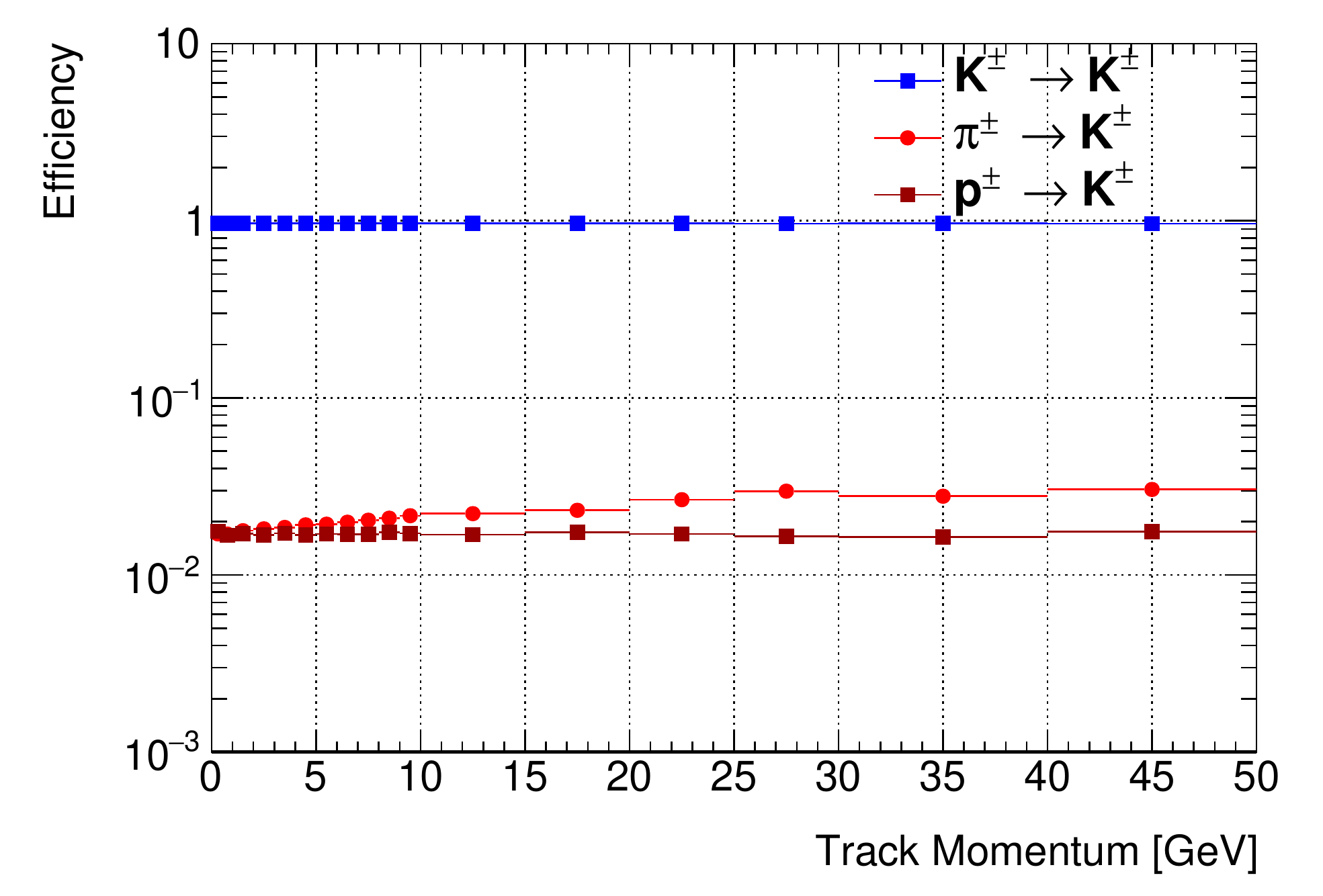}
         \caption{Kaon Identification.} 
         \label{fig:kid}
     \end{subfigure}
     \\
     \begin{subfigure}[b]{0.475\textwidth}
         \centering
         \includegraphics[width=\textwidth]{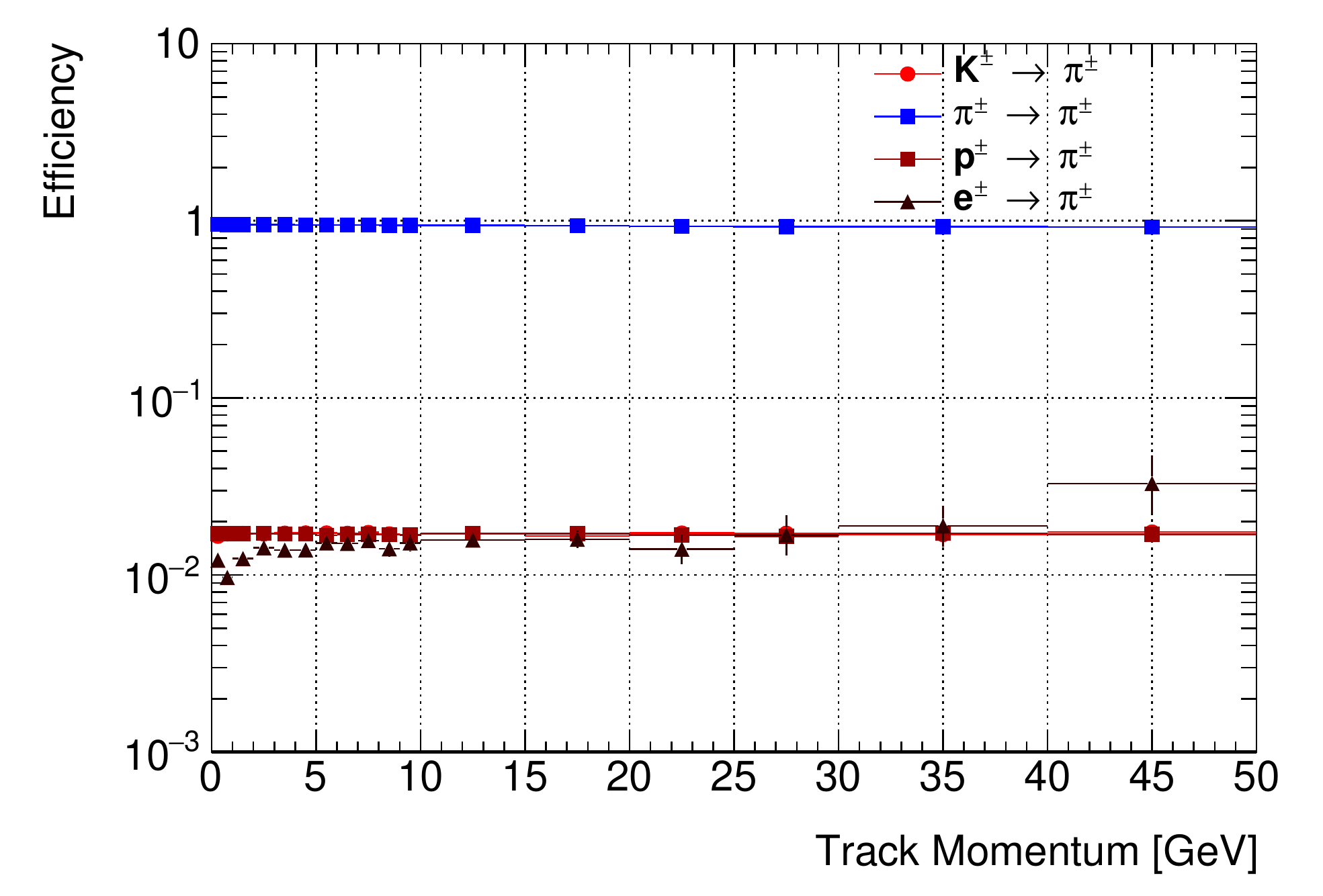}
         \caption{Pion Identification.} 
         \label{fig:pid}
     \end{subfigure}
     \hfill
     \begin{subfigure}[b]{0.475\textwidth}
         \centering
         \includegraphics[width=\textwidth]{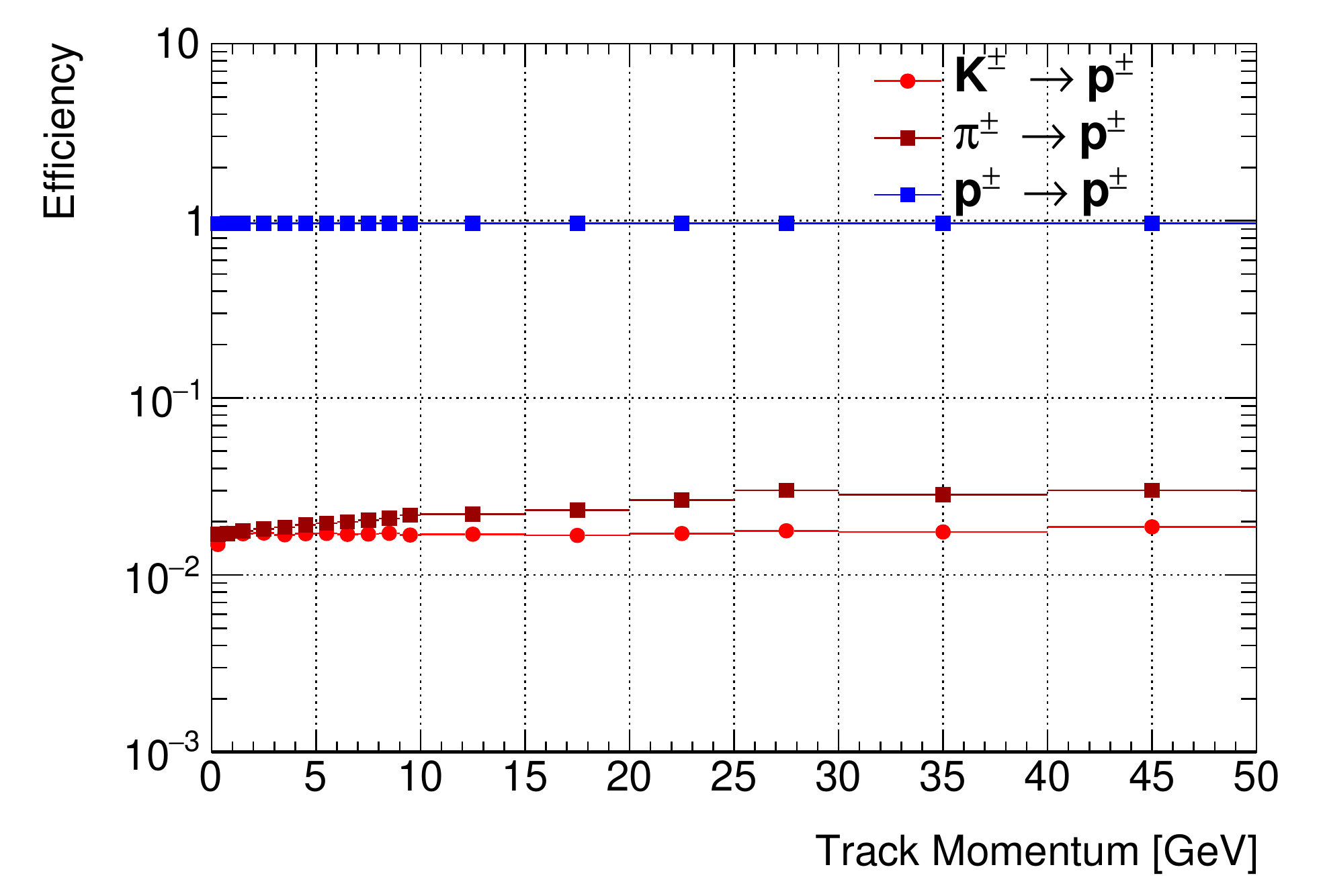}
         \caption{Proton Identification.} 
         \label{fig:prid}
     \end{subfigure}

    \caption{Particle-identification efficiency vs. momentum for all four major long-lived charged-particle species. Efficiency is determined for tracks with $|\eta|<3.5$.}
    \label{fig:particle_id}
\end{figure}

We implement a conservative, minimum particle identification efficiency map consistent with the EIC Yellow Report design parameters. The main, long-lived charged-particle species are $e^{\pm}$, $\pi^{\pm}$, $K^{\pm}$, and $p^{\pm}$. We implement $3\sigma$ separation between $e$ and $\pi$; and $3\sigma$ separation, applied in a pair-wise fashion, between $K$, $p$, and $\pi$. Example efficiency curves for track momenta up to $50~\GeV$ are shown in Fig.~\ref{fig:particle_id}.

\FloatBarrier
\section{Conclusions}
We have implemented the EIC Yellow Report detector parameters into Delphes and shown a few examples of the kind of studies that can be accomplished with it. This powerful tool could be used to complement detailed detector simulations based on Geant-4. For example, the parametrizations for tracking, calorimeter, and PID performance could be updated to parameters extracted from full simulations to provide a fast tool in which new observables can be tested in various channels (such as NC DIS, CC DIS, photoproduction, etc). We will show some examples in future work. 

\appendix

\acknowledgments

M.A. acknowledges support through DOE Contract No. DE-AC05-06OR23177 under which Jefferson Science Associates, LLC operates the Thomas Jefferson National Accelerator Facility. The work of M.A was supported by the University of California, Office of the President , MRPI award number 00010100. S.S. acknowledges support through US DOE grant DE-SC0010129. 
We gratefully acknowledge SMU's Center for Research Computation for their support and for the use of the SMU ManeFrame~II high-performance computing cluster, which enabled a portion of the simulation and analysis work in this paper.


\end{document}